\documentclass[journal, transmag]{IEEEtran}
\usepackage{amsmath,amsfonts,amssymb}
\usepackage{algorithmic}
\usepackage{array}
\usepackage{textcomp}
\usepackage{stfloats}
\usepackage{url}
\usepackage{xcolor}
\usepackage{verbatim}
\usepackage{graphicx}
\usepackage{adjustbox}
\usepackage{tikz}
\usepackage{bm}
\usepackage{multirow}
\usepackage{booktabs}
\usetikzlibrary{shapes, arrows, positioning, math, , decorations.pathreplacing}
\tikzstyle{arrow} = [thick,->,>=stealth]
\usetikzlibrary{chains,shapes.multipart}
\usetikzlibrary{shapes,calc}
\usetikzlibrary{automata}

\usepackage{hyperref}
\hypersetup{hidelinks}

\usepackage{life-hts}

\usepackage[normalem]{ulem} 

\def\BibTeX{{\rm B\kern-.05em{\sc i\kern-.025em b}\kern-.08em
    T\kern-.1667em\lower.7ex\hbox{E}\kern-.125emX}}
\usepackage{balance}
\usepackage{etoolbox}
\usepackage{eso-pic}
\makeatletter
\patchcmd{\@makecaption}
  {\scshape}
  {}
  {}
  {}
\makeatother

\begin{document}

\markboth{April 2025}
{L. Denis, E. Paakkunainen, P. Rasilo, S. Schöps, B. Vanderheyden, and C. Geuzaine: Magnetic Field Conforming Formulations for Foil Windings}

\title{\fontsize{17pt}{17pt}\selectfont Magnetic Field Conforming Formulations for Foil Windings}
\author{\IEEEauthorblockN{Louis Denis\IEEEauthorrefmark{1},
Elias Paakkunainen\IEEEauthorrefmark{2,3},
Paavo Rasilo\IEEEauthorrefmark{3},
Sebastian Schöps\IEEEauthorrefmark{2},
Benoît Vanderheyden\IEEEauthorrefmark{1}, \\
and Christophe Geuzaine\IEEEauthorrefmark{1}}
\vspace{5pt}
\IEEEauthorblockA{\IEEEauthorrefmark{1}University of Liège, 4000 Liège, Belgium\\ \IEEEauthorrefmark{2}Technical University of Darmstadt, 64289 Darmstadt, Germany \\ \IEEEauthorrefmark{3} Tampere University, 33720 Tampere, Finland}
\thanks{Corresponding author: L. Denis (e-mail: louis.denis@uliege.be).}
}

\IEEEtitleabstractindextext{%
\begin{abstract}
	We extend the foil winding homogenization method to magnetic field conforming formulations. We first propose a full magnetic field foil winding formulation by analogy with magnetic flux density conforming formulations. We then introduce the magnetic scalar potential in non-conducting regions to improve the efficiency of the model. This leads to a significant reduction in the number of degrees of freedom, particularly in 3-D applications. The proposed models are verified on two frequency-domain benchmark problems: a 2-D axisymmetric problem and a 3-D problem. They reproduce results obtained with magnetic flux density conforming formulations and with resolved conductor models that explicitly discretize all turns. Moreover, the models are applied in the transient simulation of a high-temperature superconducting coil. In all investigated configurations, the proposed models provide reliable results while considerably reducing the size of the numerical problem to be solved.
\end{abstract}

}

\AddToShipoutPicture*{
    \footnotesize\sffamily\raisebox{0.8cm}{\hspace{1.4cm}\fbox{
        \parbox{\textwidth}{
            This work has been submitted to the IEEE for possible publication. Copyright may be transferred without notice, after which this version may no longer be accessible.
            }
        }
    }
}

\maketitle
\thispagestyle{empty}
\pagestyle{empty}

\section{Introduction}
\IEEEPARstart{F}{oil windings} (FW) are used in electromagnetic devices such as transformers and inductors~\cite{Calderon-Lopez_2019aa, Das_2020aa}. Compared to conventional wire and litz windings, FW provide, e.g., higher fill factor, advantageous thermal properties~\cite{Rios_2018aa} and easier construction~\cite{Rios_2020aa}. The design of these devices requires the accurate prediction of quantities of interest, e.g. the voltage drop along the windings. The modeling of FW with the finite element (FE) method typically considers discretizing each thin conductor turn, which can be computationally expensive, particularly for 3-D applications.

Homogenization techniques~\cite{Feddi1997} aim to reduce the size of the numerical problem while preserving the accuracy of the solution, by focusing on the fields at the macroscopic scale. Such techniques have been applied to bundles of arbitrary wires in both the frequency- and time-domains~\cite{Gyselinck_2005aa, Sabariego_2008aa}. Given the specific geometry of FW and their large width-to-thickness aspect ratio, dedicated FW homogenization techniques have been proposed in \cite{DeGersem2001,Dular2002} based on magnetic flux density conforming formulations. Recently, some contributions have addressed their extension to the time-domain~\cite{Valdivieso2021} and to high frequencies~\cite{Bundschuh_2024ac}, their stability~\cite{Paakkunainen2024}, their coupling to thermal physics~\cite{Weinert2025} and their application to superconducting systems~\cite{Paakkunainen2025}. One particular advantage of the FW homogenization technique is its ability to compute the actual spatial distribution of the voltage drop along adjacent homogenized turns.

In this work, we extend the FW homogenization technique to magnetic field conforming $h$-formulations, which are derived by analogy with the FW-homogenized $a$-$v$-formulation~\cite{Dular2002}. We introduce the magnetic scalar potential in non-conducting regions to reduce the number of degrees of freedom (DoFs) required to solve the problem, leading to the $h$-$\phi$ formulation for FW. 

Besides offering significant computational gains in 3-D applications, magnetic field-conforming formulations are also beneficial for modeling superconducting systems. In particular, expressing the magnetodynamic problem in terms of the material's resistivity instead of its conductivity leads to more accurate predictions~\cite{Dular_2020aa}. Moreover, the topological similarity of FW and insulated high-temperature superconducting (HTS) coils makes the proposed models particularly relevant for the modeling of HTS applications~\cite{Paakkunainen2025}.

Section~\ref{sec:fe-formulations} describes the proposed FE formulations valid in both frequency- and time-domains. It also introduces an alternative discretization of the magnetic field, which avoids the definition of an anisotropic resistivity tensor in 3-D configurations. The developed models are verified on two benchmark problems in Section~\ref{sec:verification} and applied to the simulation of an HTS coil in Section~\ref{sec:application}.

\section{Finite Element Formulations \label{sec:fe-formulations}}
A coil of $N_{\text{c}}$ insulated thin conductors (or \textit{foils}) is considered, which constitutes the conducting subset $\Oc$ of the computational domain~$\O$, whose cross-section in the $\alpha$-$\beta$ plane is shown in Fig.~\ref{fig:sketch}. The turns being arranged in series, the net current in each conductor $\Oci$ is $I_i = \It$.

\subsection{Magnetic Flux Density Conforming Formulation}
\begin{figure}[!t]
    \centering
    \includegraphics[width=\columnwidth]{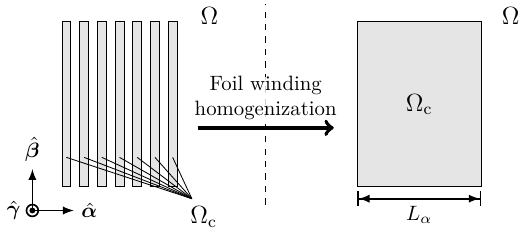}
    \caption{FW homogenization: $N_{\text{c}}$ resolved conductors (left) replaced by a single homogenized bulk of thickness $L_{\alpha}$ (right).}
    \label{fig:sketch}
\end{figure}

The magnetodynamic $a$-$v$ formulation consists in finding the gauged magnetic vector potential $\a$ and the scalar electric potential $v$ such that \cite{Dular1999aa}
\begin{equation}
\volInt{\nu\curl{\a}}{\curl{\a}'}{\O}\! + \volInt{\sigma\partial_t\a}{\a'}{\Oc}\!\! + \volInt{\sigma\grad v}{\a'}{\Oc} = 0 \label{eq:av-1}
\end{equation}
\begin{equation}
\volInt{\sigma\partial_t\a}{\grad v'}{\Oc}\!\! + \volInt{\sigma\grad v}{\grad v'}{\Oc} \!\! + \sum_{i=1}^{N_{\text{c}}} \It V_i' = 0 \label{eq:av-2}
\end{equation}
holds $\forall \a', v', V_i'$, with $\nu$~the magnetic reluctivity and $\sigma$~the electrical conductivity. Here, $\grad v = \sum_{i=1}^{\Nc} V_i\,\grad v_{\text{s},i}$ and the $\grad v_{\text{s},i}$ denote global basis functions~\cite{Dular1999aa}, or \textit{winding functions}~\cite{Schoeps2013}. Test functions are denoted $\cdot'$, whereas $\volInt{\cdot}{\cdot}{\O}$ represents the volume integral over $\O$ of the inner product of its arguments. After discretization, $\a$ is gauged using the co-tree gauge~\cite{Creuse2019} in 3-D, whereas $\a = a \cdot \hat{\bm{\gamma}}$ implicitly satisfies the Coulomb gauge in 2-D.

In~\eqref{eq:av-1}-\eqref{eq:av-2}, each turn is treated as an independent \textit{massive} conductor. The magnetic vector potential $\a$ is approximated using edge shape functions, whereas $v_{\text{s},i}$ is the sum of all nodal shape functions linked to a cross-section of turn $\Oci$ with support limited to a transition layer~\cite{Dular1999aa}. In 2-D, $\grad v_{\text{s},i} = \hat{\bm{\gamma}}$.

As shown in Fig.~\ref{fig:sketch}, the FW homogenization ansatz \cite{DeGersem2001,Dular2002} replaces the successive turns $\Oci$ of the coil by a single homogenized bulk. Consequently, the gradient of the electric scalar potential is extended to a continuum:
\begin{equation}
    \grad v = \Phi(\alpha)\,\grad v_{\text{s}} = \sum_{k=1}^{N_{\text{b}}} u_k\,p_k(\alpha)\,\grad v_{\text{s}}, \label{eq:fw-ansatz}
\end{equation}
in which the $N_{\text{b}}$ spatially dependent global shape functions $p_k(\alpha)$ are not necessarily mesh related. The weak imposition of the current\,\eqref{eq:av-2} becomes
\begin{multline}
    \volInt{\sigma\partial_t\a}{\Phi' \grad v_{\text{s}}}{\Oc} + \volInt{\sigma\,\Phi\,\grad v_{\text{s}}}{\Phi'\grad v_{\text{s}}}{\Oc}  \\ + \frac{N_{\text{c}}}{L_{\alpha}}\int_{L_{\alpha}} \It\,\Phi'\,d\alpha = 0, \label{eq:av-2-foil}
\end{multline}
using $\grad v' = \Phi'\,\grad v_{\text{s}}$, with the single $v_{\text{s}}$ defined as the sum of all nodal shape functions on one cross-section of the coil~\cite{Dular2002}. The definition of an effective anisotropic conductivity $\bm{\sigma}$, combined with~\eqref{eq:av-2-foil}, enforces the current conservation in every (virtual) elementary foil inside $\Oc$ while allowing skin effect along the width ($\beta$-axis) of the foils \cite{Dular2002}. Recently, this $a$-$v$ foil winding model has been extended to the $j$-$a$-$v$ formulation in \cite{Paakkunainen2025}.

\subsection{Magnetic Field Conforming Formulations}
The classical full-$h$ formulation, relying on winding functions \cite{Schoeps2013} for the weak imposition of the current, consists in finding the magnetic field $\h$ and $v$ such that
\begin{multline}
    \volInt{\mu\partial_t\h}{\h'}{\O} + \volInt{\rho\,\curl{\h}}{\curl{\h}'}{\O} \\ + \volInt{\grad v}{\curl{\h}'}{\Oc} = 0 \label{eq:htot-1}
\end{multline}
\begin{equation}
    \volInt{\curl \h}{\grad v'}{\Oc} - \sum_{i=1}^{N_{\text{c}}} \It V_i' = 0 \label{eq:htot-2}
\end{equation}
holds $\forall \h', v', V_i'$, with $\mu$ the magnetic permeability and $\rho$~the electrical resistivity. Here, $\grad v =\sum_{i=1}^{\Nc} V_i\,\grad v_{\text{s},i}$ as successive turns are treated as \textit{massive} conductors.

By analogy with the $a$-$v$ formulation and considering the FW homogenization ansatz~\eqref{eq:fw-ansatz}, we replace~\eqref{eq:htot-2} by
\begin{equation}
    \volInt{\curl \h}{\Phi' \grad v_{\text{s}}}{\Oc} - \frac{N_{\text{c}}}{L_{\alpha}}\int_{L_{\alpha}} \It\,\Phi'\,d\alpha = 0. \label{eq:htot-2-foil}
\end{equation}
The combination of~\eqref{eq:fw-ansatz},~\eqref{eq:htot-1} and~\eqref{eq:htot-2-foil} leads to the full-$h$ \textit{FW formulation}. At the discrete level, $\h$ is approximated using edge shape functions, exclusively. One drawback of the full-$h$ formulation is the need to define a spurious resistivity in the non-conducting domain $\Occ$, which is for example set to $\rho=1$~$\Omega$\,m~\cite{Dlotko2019}.

To reduce the number of required DoFs to solve the problem, we extend the proposed model to the $h$-$\phi$ formulation. In the non-conducting domain $\Occ = \O \setminus \Oc$, 
\begin{equation}
\curl \h = \vec 0  \quad \text{such that} \quad \h = - \grad \phi, \label{eq:h-phi-def}
\end{equation} 
with $\phi$ the magnetic scalar potential. By analogy with the classical $h$-$\phi$ formulation \cite{Dular1999ab}, we discretize $\h$ as
\begin{equation}
    \h = \sum_{e \in \mathcal{E}(\Oc \setminus \partial \Oc)} h_e\,\vec\psi_e + \sum_{n \in \mathcal{N}(\Occ)} \phi_n\,\grad\psi_n + \If \,\vec c_{\text{f}} \label{eq:h-discretization-foil}
\end{equation}
with $\vec \psi_e$ (resp. $\psi_n$) edge (resp. nodal) shape functions and $\vec c_{\text{f}}$ a single global cut basis function associated to the homogenized bulk $\Oc$, which allows a net circulation of the magnetic field around the coil. Notably, a single cut must be defined using the proposed model, by contrast to \cite{Dular1999ab} which considers one cut per turn. The \textit{cut} corresponds to a global edge cohomology basis function dual to a closed loop around the bulk, as described in~\cite{Pellikka_2013aa}.

Instead of imposing the current constraint by strongly fixing the associated global DoF~$\If$ in~\eqref{eq:h-discretization-foil}, the current is weakly enforced using~\eqref{eq:htot-2-foil}. This approach corresponds to the weak current imposition via Lagrange multipliers~\cite{Babuska2003}. The global DoF~$\If$ will be equal, via~\eqref{eq:htot-2-foil}, to the total transport current through the coil cross-section: $\If = N_{\text{c}}\It$. The combination of~\eqref{eq:fw-ansatz},~\eqref{eq:htot-1} and~\eqref{eq:htot-2-foil}, together with~\eqref{eq:h-discretization-foil}, leads to the \textit{$h$-$\phi$ FW formulation}.

The second term in~\eqref{eq:htot-1} simplifies to 
\begin{equation}
    \volInt{\rho\,\curl{\h}}{\curl{\h}'}{\O} = \volInt{\rho\,\curl{\h}}{\curl{\h}'}{\Oc} \label{eq:hphi-1}
\end{equation}
by definition of~\eqref{eq:h-phi-def}, which does not require the definition of a spurious resistivity in the non-conducting domain. In 2-D configurations, the current density $\j = \curl \h$ is oriented along the  $\gamma$-axis such that the effective resistivity $\rho$ in~\eqref{eq:hphi-1} can be modelled by the scalar $\rho_0$, the resistivity of individual foils divided by their fill factor. In 3-D, it is necessary to define an anisotropic resistivity tensor $\bm{\rho}$ to avoid spurious currents between adjacent virtual elementary foils. It is defined in local $\alpha$-$\beta$-$\gamma$ coordinates:
\begin{equation}
    \bm{\rho} = \begin{pmatrix}
        \rho_{\alpha} & 0 & 0 \\
        0 & \rho_{\beta} & 0 \\
        0 & 0 & \rho_{\gamma}
    \end{pmatrix} = \begin{pmatrix}
        r_{\text{a}} \cdot \rho_0 & 0 & 0 \\
        0 & \rho_0 & 0 \\
        0 & 0 & \rho_0
    \end{pmatrix}, \label{eq:anisotropy}
\end{equation}
with $r_{\text{a}}$ the anisotropy ratio, which should ideally have an infinite value. However, excessively large values of $r_{\text{a}}$ can lead to very ill-conditioned FE systems. In practice, a trade-off value of $r_{\text{a}} = 10^4$ is found to be sufficient to enforce current conservation in each virtual foil. When the orientation of the foils varies with space, care should be taken to integrate~\eqref{eq:hphi-1} appropriately: selective reduced integration~\cite{Malkus1978} is used by integrating~\eqref{eq:hphi-1} with a single Gauss point. This also requires a structured hexahedral mesh in the homogenized bulk.

As an alternative solution to the anisotropic resistivity tensor~\eqref{eq:anisotropy}, the $\h$-discretization~\eqref{eq:h-discretization-foil} can be adapted to avoid spurious current flow in the $\hat{\vec \alpha}$-direction normal to the foils by construction. A structured hexahedral mesh of the homogenized bulk is constructed, which is composed of radial layers of elements aligned with the $\gamma$-axis that are not required to coincide with the actual discretized foils.

\begin{figure}[!t]
    \centering
    \includegraphics[width=0.9\columnwidth]{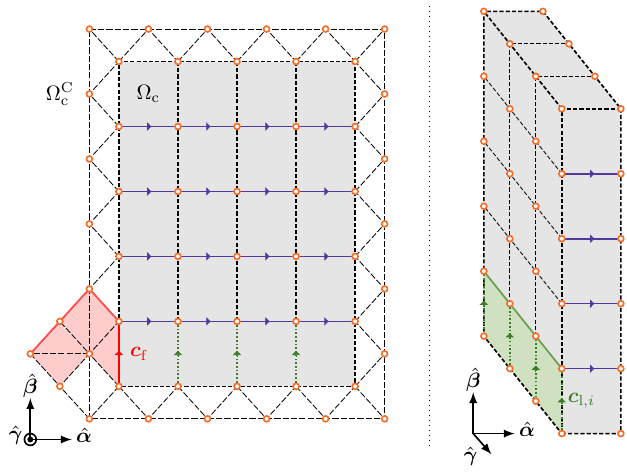}
    \caption{Illustration of entities involved in the discretization of the $t$-$\omega$ FW formulation, visualized on a structured hexahedral mesh of the homogenized bulk~$\Oc$. Left: cross-section of the homogenized bulk with the global cut basis function $\bm{c}_{\text{f}}$ and its support. Right: single layer $i$ of elements in the radial $\hat{\vec\alpha}$-direction with the layer cut-like basis function $\bm{c}_{\text{l},i}$ and its support. Nodal DoFs are in orange, remaining edge DoFs are in blue.}
    \label{fig:tw-function_space}
\end{figure}

The adapted discretization is visualized on the mesh represented in Fig.~\ref{fig:tw-function_space}. As shown, it takes into account DoFs related to edges only aligned with~$\hat{\vec \alpha}$, denoted by $h_{e,\alpha}$, to prevent current flow between successive radial layers. Accordingly, the magnetic scalar potential, discretized with nodal DoFs, is extended to the homogenized bulk to retrieve the magnetic field component orthogonal to $\hat{\vec \alpha}$. To allow for a net circulation of the magnetic field between successive layers, one cut-like basis function $\vec c_{\text{l},i}$ is introduced per layer, as represented in Fig.~\ref{fig:tw-function_space}. It is defined as the sum of all edge shape functions aligned with $\hat{\vec \beta}$, perpendicular to one arbitrary consecutive set of edges along $\hat{\vec \gamma}$. Outside of the bulk, one single global cut function $\vec c_{\text{f}}$ is still considered. Finally, the alternative discretization of the magnetic field for 3-D simulations reads
\begin{multline}
    \h = \sum_{e,\alpha \in \mathcal{E}_\alpha(\Oc \setminus \partial \Oc)} h_{e,\alpha}\,\vec\psi_{e,\alpha} + \sum_{n \in \mathcal{N}(\O)} \phi_n\,\grad\psi_n \\ + \If \,\vec c_{\text{f}} + \sum_{i=2}^{N_{\text{l}}} I_{{\text{l}},i}\,\vec c_{\text{l},i}, \label{eq:tw-discretization-foil}
\end{multline}
with $N_{\text{l}}$ the number of radial element layers in the mesh. This discretization shares common features with the $t$-$\omega$ formulation~\cite{Pellikka_2013aa}, considering the electric vector potential $\vec t = t_{\alpha} \cdot \hat{\vec \alpha}$ aligned with the radial direction as in the homogenized $t$-$a$ formulation~\cite{VargasLlanos2022}. To avoid confusion with the $h$-$\phi$ FW formulation, we refer to the combination of~\eqref{eq:fw-ansatz},~\eqref{eq:htot-1} and~\eqref{eq:htot-2-foil}, together with~\eqref{eq:tw-discretization-foil}, as the \textit{$t$-$\omega$ FW formulation}. Considering~\eqref{eq:tw-discretization-foil}, the isotropic effective resistivity $\rho$ in~\eqref{eq:hphi-1} simply corresponds to $\rho_0$, as radial current sharing between turns is avoided by construction.

Advantages of the $t$-$\omega$ FW formulation include fewer DoFs compared to the $h$-$\phi$ FW formulation and an isotropic resistivity in the homogenized bulk. On the other hand, the $t$-$\omega$ discretization~\eqref{eq:tw-discretization-foil} is more complex to implement than~\eqref{eq:h-discretization-foil} using commercial FE software, as it requires the definition of several cut-like basis functions within the structured mesh of the homogenized bulk.

\section{Numerical Verification \label{sec:verification}}
The FE formulations described in the previous section are implemented in the open-source software GetDP~\cite{Dular_1998ac} and meshes are generated using Gmsh~\cite{Geuzaine_2009ab}. Models presented in this section are available online~\cite{MyZenodo}.

The proposed magnetic field conforming FW formulations are verified on two frequency-domain simulations: a 2-D axisymmetric problem and a 3-D problem. The results are compared to magnetic flux density conforming FW formulations and to resolved-conductor models that explicitly discretize all turns.
\subsection{2-D Axisymmetric Problem}
\begin{figure}[t!]
    \centering
    \includegraphics[width=0.7\columnwidth]{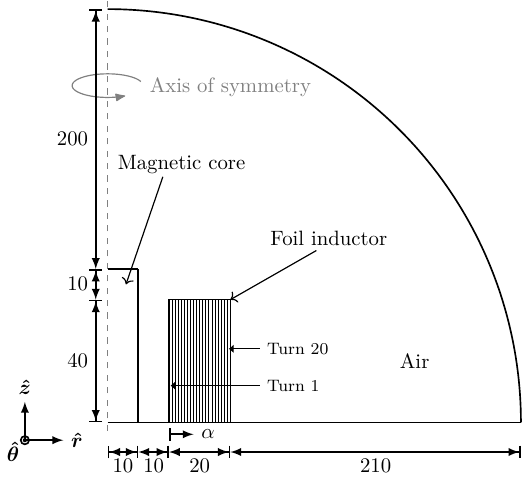}
    \caption{Geometry (not to scale) of the 2-D axisymmetric verification problem: 20-turn foil winding inductor around a magnetic core ($\mu_{\text{r}} = 10$), both placed inside air. Units: mm.}
    \label{fig:geom2D}
\end{figure}

In the first verification problem, we examine a foil winding around a magnetic core in a 2-D axisymmetric setting as represented in Fig.~\ref{fig:geom2D}. The foil winding has 20 turns with material conductivity $5.9\times10^7$~$\Omega^{-1}$\,m$^{-1}$, and the winding is driven with a current of $\It=1$~A at $50$~Hz. At the considered frequency, capacitive effects can be neglected~\cite{Bundschuh_2024ac}. Eddy currents in the open core are also neglected, such that the union of the core and the air constitute $\Occ$. The winding is assumed to have unit fill factor.

\begin{figure}[t!]
    \centering
    \includegraphics{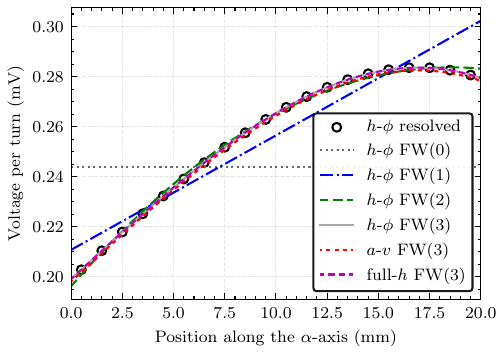}
    \caption{Voltage per turn in the 20-foil winding computed with the $h$-$\phi$ resolved, $h$-$\phi$ FW, full-$h$ FW and $a$-$v$ FW models. The FW model approximates the voltage with global polynomials of order $p=N_{\text{b}}-1 \in \{0,1,2,3\}$. The same mesh is used for all models.}
    \label{fig:voltage2D}
\end{figure}

The proposed full-$h$ and $h$-$\phi$ FW formulations are compared with the $a$-$v$ FW formulation and the $h$-$\phi$ formulated resolved model. The global voltage function $\Phi(\alpha)$ is discretized using polynomials with global support. However, other choices for the basis functions are also possible, e.g., B-splines with local support~\cite{Bundschuh_2024ac}. Fig.~\ref{fig:voltage2D} shows the voltage per turn for the resolved model and the voltage distribution for the homogenized models. The models show good agreement, provided that the polynomial order for $\Phi(\alpha)$ is sufficiently high. With the $h$-$\phi$ FW formulation, a second-order polynomial for $\Phi(\alpha)$ already provides satisfying results, while a third-order polynomial allows to reproduce the voltage values obtained with the resolved model, with only four degrees of freedom associated to the voltage continuum instead of 20. Similar results are obtained with the other FW formulations.

\begin{figure}[t!]
    \centering
    \includegraphics{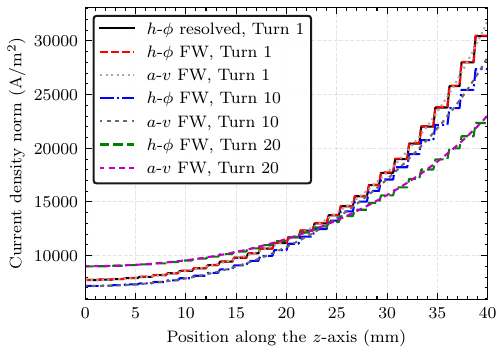}
    \caption{Current density distribution along the center of different (virtual) turns of the FW, computed with the various models. FW models consider a third-order global polynomial for $\Phi(\alpha)$. The same mesh is used for all models.}
    \label{fig:j2D}
\end{figure}

The current density distributions computed with the different formulations are compared in Fig.~\ref{fig:j2D}. As may be observed, the different models are in good agreement, further verifying the proposed formulations. This highlights the effective current conservation in each virtual foil through the FW weak formulation. Notably, the current density computed with $h$-based FW formulations is constant per element, as expected when using first-order basis functions for the $h$-discretization.

\subsection{3-D Problem}
\begin{figure}[t!]
    \centering
    \includegraphics[width=0.9\columnwidth]{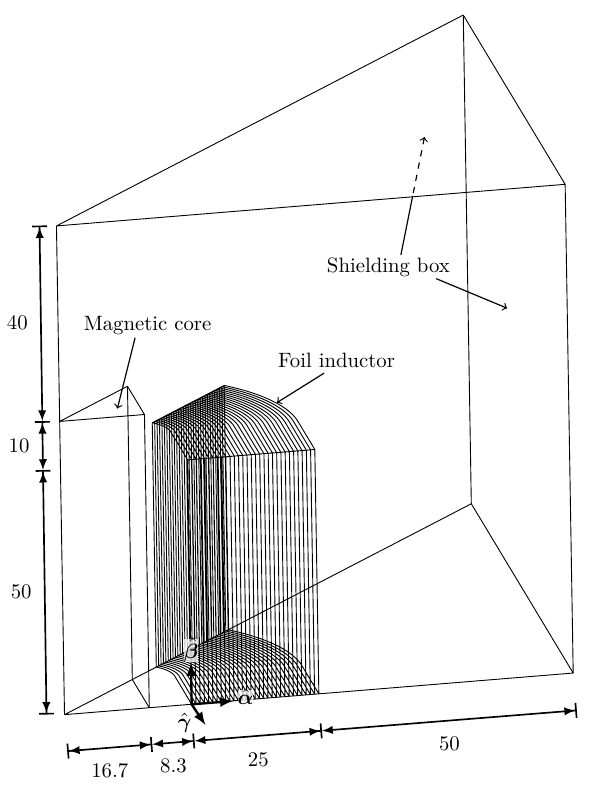}
    \caption{Foil winding inductor geometry for the 3-D verification problem: 30 foil conductors are wound around an open magnetic core ($\mu_{\text{r}} = 10$), with a highly magnetic shielding box covering the structure. Figure adapted from~\cite{Dular2002}. Units: mm.}
    \label{fig:sketch3D}
\end{figure}

The second verification problem consists of the 3-D magnetodynamic problem studied in~\cite{Dular2002}, whose geometry is shown in Fig.~\ref{fig:sketch3D}, with 1/16 of the structure modeled by symmetry. The inductor is constituted of $\Nc=30$ insulated foil conductors of conductivity $1.475\times10^7$~$\Omega^{-1}$\,m$^{-1}$ and with unit fill factor, for simplicity. The highly magnetic shielding box is assumed to ensure a purely normal magnetic flux density on its inner boundary. The coil is driven by an ac current at 50~Hz and its net magnetomotive force amplitude is $12$~A ($I_{\text{t}}=0.2$~A per half-turn).

\begin{figure}[t!]
    \centering
    \includegraphics{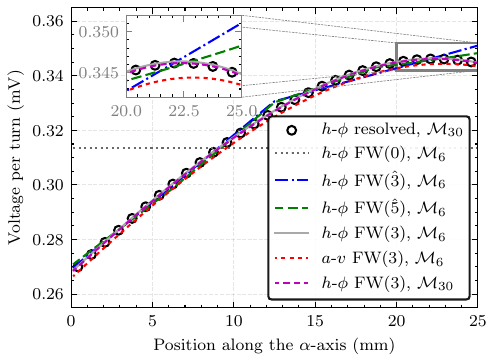}
    \caption{Voltage per turn in the 30-foil winding computed with the $h$-$\phi$ resolved, $h$-$\phi$ FW and $a$-$v$ FW models. The FW model approximates the voltage with global polynomials of order $N_{\text{b}}-1 \in \{0,3\}$, or with a set of $N_{\text{b}} \in \{3,5\}$ piecewise linear functions denoted by FW($\hat{N_{\text{b}}}$). Two meshes in the homogenized bulk are considered: $\mathcal{M}_{30}$ and $\mathcal{M}_6$ with respectively 30 and 6 layers of elements along the bulk's radius.}
    \label{fig:voltage3D}
\end{figure}

The problem is first solved considering the foil conductors as independent massive conductors using the resolved $h$-$\phi$ formulation~\cite{Dular1999ab}. The corresponding voltage distribution (constant per conductor) is represented in Fig.~\ref{fig:voltage3D}. As shown in Fig.~\ref{fig:voltage3D}, a third-order global polynomial for the $\Phi(\alpha)$ discretization again provides good agreement between the $h$-$\phi$ FW model and the resolved model. Also, Fig.~\ref{fig:voltage3D} shows results with piecewise linear functions for the voltage distribution, which agree well with other discretizations, but are less accurate than the third-order global polynomial for $\Phi(\alpha)$.

Notably, a structured mesh $\mathcal{M}_{6}$ constituted of only 6 layers of hexahedral elements along the radial direction of the homogenized bulk (total $h$-$\phi$ FW \#DoFs = 36097) produces results close to those obtained with the structured mesh $\mathcal{M}_{30}$ of 30 layers used to also solve the resolved model (total $h$-$\phi$ FW \#DoFs = 95346). This highlights the computational advantage of the homogenization technique itself, which does not require all turns to be discretized.

The full-$h$ FW and $t$-$\omega$ FW results are not shown in Fig.~\ref{fig:voltage3D} since the corresponding voltage curves are almost identical to the $h$-$\phi$ FW results. This is verified by computing the coefficient of determination $R^2$ obtained with the voltage distributions $\Phi$ computed with respect to the $h$-$\phi$ FW distribution $\Phi_{h\phi}$ (mean value: $\bar{\Phi}_{h\phi}$) as
\begin{equation}
    R^2 = 1 - \frac{\int_{L_{\alpha}} (\Phi_{h\phi} - \Phi)^2\,d\alpha}{\int_{L_{\alpha}} (\Phi_{h\phi} - \bar{\Phi}_{h\phi})^2\,d\alpha}.
\end{equation}
For the $t$-$\omega$ FW and full-$h$ FW models, using a third-order global polynomial for $\Phi(\alpha)$ and mesh $\mathcal{M}_6$, $1-R^2$ is equal to $10^{-8}$ and $2\times10^{-10}$, respectively. This indicates that the $t$-$\omega$ FW and full-$h$ FW models are equivalent to the $h$-$\phi$ FW model in terms of accuracy, as expected. Moreover, the $h$-based FW models reproduce the results obtained with the $a$-$v$ FW model~\cite{Dular2002}, since its $R^2$ coefficient is $0.992$.

\begin{figure}[t!]
    \centering
    \includegraphics{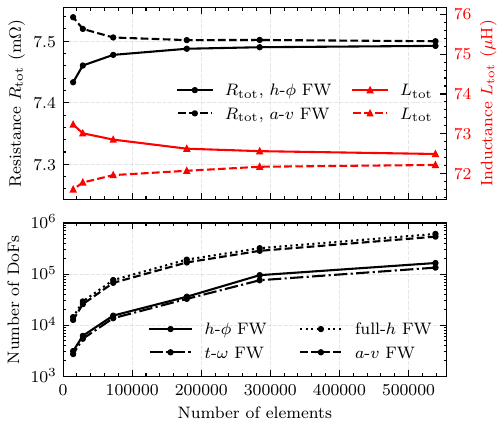}
    \caption{Top: resistance and inductance computed with the $h$-$\phi$ FW and $a$-$v$ FW models as a function of mesh refinement. The voltage $\Phi(\alpha)$ is approximated with a third-order global polynomial. Bottom: corresponding number of DoFs for the $h$-$\phi$ FW, $t$-$\omega$ FW, full-$h$ FW and $a$-$v$ FW models.
    }
    \label{fig:convergence3D}
\end{figure}

The convergence of global quantities (total resistance~$R_{\text{tot}}$ and total inductance $L_{\text{tot}}$) computed with the FW models is shown in Fig.~\ref{fig:convergence3D}. These quantities are deduced from the computation of the total voltage drop $V_{\text{tot}}$ across the FW~\cite{Dular2002}:
\begin{equation}
    V_{\text{tot}} = \frac{\Nc}{L_{\alpha}} \int_{L_{\alpha}} \Phi(\alpha) \,d\alpha = (R_{\text{tot}} + j2\pi f L_{\text{tot}}) I_{\text{t}},
\end{equation}
with $j$ the imaginary unit. Fig.~\ref{fig:convergence3D} highlights the duality of the $a$-$v$ and $h$-$\phi$ formulations~\cite{Rikabi1988}, as $R_{\text{tot}}$ and $L_{\text{tot}}$ converge in a complementary manner with mesh refinement, which verifies the consistency of the proposed model. Again, the results obtained with the full-$h$ FW and $t$-$\omega$ FW models are not shown in Fig.~\ref{fig:convergence3D} as they coincide with the $h$-$\phi$ FW results. The corresponding number of DoFs is also represented in Fig.~\ref{fig:convergence3D}. The $h$-$\phi$ FW and $t$-$\omega$ FW models require much fewer DoFs than the $a$-$v$ FW model, with a respective reduction of 69.5\% and 75.2\% in the number of DoFs for the densest mesh, the $t$-$\omega$ FW model being the most economical. On the other hand, the full-$h$ FW model requires more DoFs than the $a$-$v$ FW model (13\% more DoFs for the densest mesh). This emphasizes the large computational gain obtained with the $h$-$\phi$ FW and $t$-$\omega$ FW models in 3-D configurations.

\begin{figure}[t!]
    \centering
    \includegraphics{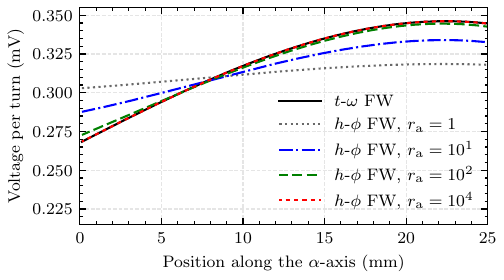}
    \caption{Voltage curves obtained with the $t$-$\omega$ FW and $h$-$\phi$ FW models with various anisotropy ratios $r_{\text{a}}$ in~\eqref{eq:anisotropy}, using a third-order global polynomial for $\Phi(\alpha)$ and mesh $\mathcal{M}_{30}$.}
    \label{fig:anisotropy3D}
\end{figure}

The influence of the anisotropy ratio in~\eqref{eq:anisotropy} is studied in Fig.~\ref{fig:anisotropy3D}. The solid line represents the voltage curve obtained with the $t$-$\omega$ FW model and its isotropic resistivity, as radial current sharing between turns is avoided by construction of~\eqref{eq:tw-discretization-foil}. As highlighted in Fig.~\ref{fig:anisotropy3D}, the $h$-$\phi$ FW results (dashed lines) converge towards the $t$-$\omega$ FW model when the anisotropy ratio increases, and $r_{\text{a}}=10^4$ effectively prevents current sharing between virtual foils. This verifies the equivalence between the $t$-$\omega$ FW and $h$-$\phi$ FW models when the anisotropy ratio is sufficiently high.

\section{2-D HTS Coil Simulation  \label{sec:application}}
\begin{figure}[t!]
    \centering
    \includegraphics{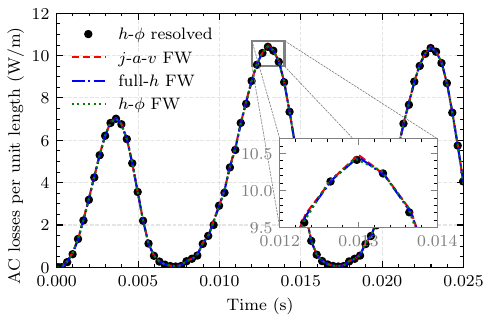}
    \caption{
        Evolution of AC losses per unit length in the HTS coil~\cite{Paakkunainen2025} against time obtained with various FW formulations using a third-order global polynomial for $\Phi(\alpha)$, compared to results from the resolved $h$-$\phi$ formulation~\cite{Dular1999ab}.
    }
    \label{fig:results_hts}
\end{figure}

To highlight the robustness of the proposed formulations, a strongly non-linear time-dependent problem is considered, namely the 2-D simulation of an insulated HTS coil made of 20 turns described in~\cite{Paakkunainen2025}, with fill factor 0.01. Numerical results, obtained with a third-order global polynomial for $\Phi(\alpha)$, are compared to the $j$-$a$-$v$ FW model~\cite{Paakkunainen2025} and to the resolved $h$-$\phi$ formulation~\cite{Dular1999ab} in Fig.~\ref{fig:results_hts}. As highlighted, the AC losses are accurately reproduced with the two $h$-conforming FW formulations. The number of DoFs required to solve the problem is 8.7k, 14.3k, 6k for the $j$-$a$-$v$ FW, full-$h$ FW and $h$-$\phi$ FW formulations, respectively. Once again, the $h$-$\phi$ FW model enables a significant reduction in problem size.

\section{Conclusion \label{sec:conclusion}}
In this paper, magnetic field conforming formulations are proposed for the homogenization of FWs, both in 2-D and 3-D. These models are implemented using open-source software and have been verified against different formulations on two benchmark problems. In the 2-D axisymmetric configuration, as well as in the 3-D configuration, the full-$h$ FW and $h$-$\phi$ FW homogenized models reproduce results obtained with the $a$-$v$ FW model and the resolved $h$-$\phi$ model. Whereas the 2-D models support arbitrary meshes, the $h$-based 3-D models are limited to structured hexahedral meshes in the homogenized coil, while they also require the definition of an anisotropic resistivity tensor to prevent spurious current sharing between adjacent turns. A significant advantage of the $h$-$\phi$ FW model is the single cut function associated to the homogenized bulk, as effective current conservation in each virtual foil is ensured in the weak formulation. This differs from the resolved model which considers one cut function per turn. An alternative discretization of the magnetic field is also introduced, corresponding to the so-called $t$-$\omega$ FW model. It allows an isotropic resistivity to be used since spurious radial current sharing is avoided by construction of the function space. It matches the results from the full-$h$ FW and $h$-$\phi$ FW models.

As expected, no computational gain is obtained with the full-$h$ FW model compared to the $a$-$v$ FW model. However, the $h$-$\phi$ FW and $t$-$\omega$ FW models significantly reduce the number of DoFs required to solve the problem. This is particularly striking in 3-D configurations as the size of the problem is reduced by up to 75\% with the proposed models relative to the $a$-$v$ FW model. While the $t$-$\omega$ FW model is slightly more efficient, the $h$-$\phi$ FW model is easier to implement as it requires only one single global cut function associated to the homogenized foil winding.

Moreover, the proposed models are not restricted to frequency-domain simulations, as they are also applied to the non-linear transient simulation of an HTS coil, again demonstrating reliable results with a reduced number of DoFs. This highlights the robustness of the developed open-source models, which constitute an efficient solution to simulate large-scale 3-D FW inductors.

\section*{Acknowledgment}
L. Denis is a research fellow funded by the F.R.S-FNRS. The work of E. Paakkunainen is supported by the Graduate School CE within Computational Engineering at the Technical University of Darmstadt.

The authors would like to thank E. Schnaubelt from CERN for sharing his insights on winding functions and their use in the $h$-formulation, F. Henrotte from University of Liège and J. Dular from CERN for the fruitful discussions on the topic.

\end{document}